\documentclass[aps,prl,amssymb,twocolumn,superscriptaddress,showpacs]{revtex4}

\usepackage{graphicx}
\usepackage{dcolumn}
\usepackage{bm}

\begin{document}

\newcommand{\beq}{\begin{equation}}
\newcommand{\eeq}{\end{equation}}
\newcommand{\barr}{\begin{eqnarray}}
\newcommand{\earr}{\end{eqnarray}}

\title{Direct experimental evidence of free fermion antibunching}
\author{M. Iannuzzi}
\affiliation{Dipartimento di Fisica, Universit\`a di Roma ``Tor
Vergata", I-00133 Roma, Italy}
\author{A. Orecchini}
\affiliation{Dipartimento di Fisica, Universit\`a di Perugia,
I-06123 Perugia, Italy} \affiliation{INFM, Perugia, I-06123
Perugia, Italy}
\author{F. Sacchetti}
\affiliation{Dipartimento di Fisica, Universit\`a di Perugia,
I-06123 Perugia, Italy} \affiliation{INFM, Perugia, I-06123
Perugia, Italy}
\author{P. Facchi}
\affiliation{Dipartimento di Matematica, Universit\`a di Bari,
        I-70125  Bari, Italy}
\affiliation{INFN, Sezione di Bari, I-70126 Bari, Italy}
\author{S. Pascazio} \affiliation{Dipartimento di Fisica,
Universit\`a di Bari,
        I-70126  Bari, Italy}
\affiliation{INFN, Sezione di Bari, I-70126 Bari, Italy}

\date{\today}

\begin{abstract}
Fermion antibunching was observed on a beam of free noninteracting
neutrons. A monochromatic beam of thermal neutrons was first split
by a graphite single crystal, then fed to two detectors,
displaying a reduced coincidence rate. The result is a fermionic
complement to the Hanbury Brown and Twiss effect for photons.
\end{abstract}


\maketitle

Over the past three decades, the research on the foundations of
quantum mechanics has been enriched by many experiments on thermal
neutrons, in particular by several enlightening results about the
coherence properties and the physical nature of the wave function
describing the behavior of a massive particle \cite{gen}. A
general property of fermions is that of being characterized by an
antisymmetric wave function: the second-order correlation function
of a fermion gas exhibits an anticorrelation in the intensity
fluctuations, in particular interference in the coincidence
distributions of identical particles. The present Letter describes
a new contribution in this field: an experiment on thermal
neutrons that brings to light the fermion antibunching effect in a
beam of free noninteracting particles. A monochromatic beam of
thermal neutrons was first split by a graphite single crystal,
then fed to two detectors, displaying a reduced coincidence rate.
The result is a fermionic complement to the seminal Hanbury Brown
and Twiss effect for bosons (photons)\cite{sei}.

The consequences of antisymmetry are well known in condensed
matter physics, where the electronic states display a strong
quantum entanglement and are confined within the Fermi surface.
Interesting experiments with electron beams have confirmed these
effects \cite{electron1,electron2,electron3}. In the case of
almost free particles, an anticorrelation was observed in the
coincidence spectrum of neutrons from compound-nuclear reaction at
small relative momentum \cite{tre,qua}. However, such a physical
system is not a good representative sample of a statistical
ensemble of non-interacting identical fermions. A monochromatic
beam of thermal neutrons from a nuclear reactor represents much
better a statistical ensemble of free particles. Nevertheless, the
observation of thermal-neutron antibunching by means of
coincidence measurements on such beams with the available
instrumentation did not appear to be feasible up to now, mainly
because the mean number of fermions obtainable per unit cell of
phase-space, to which the signal-to-noise ratio is proportional,
was so low that a measurement time of several years was estimated
\cite{cin}.

We shall show below that, with present-day available advanced
instrumentation, a very accurately designed setup and a precise
knowledge of the statistical properties of the neutron source, the
experiment is feasible. In this Letter we shall describe some
measurements carried out at the Institute Laue Langevin, Grenoble,
France.

How can one directly bring to light an anticorrelation effect in a
neutron beam? In a gas of fermions there is a certain tendency for
particles of the same spin to avoid each other, a tendency arising
from the exchange antisymmetry of the wave function: two fermions
in the same spin state cannot occupy at the same time the same
point in space, and therefore the probability amplitude for their
being close together must be small. We just want to observe such
an effect in a beam of thermal neutrons.

\begin{figure}
\begin{center}
\includegraphics[width=0.9\linewidth]{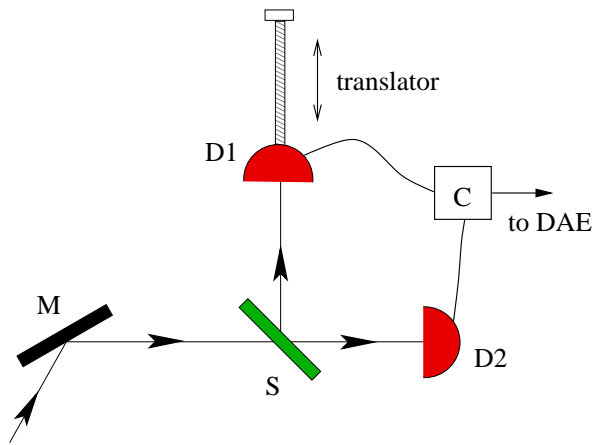}
\end{center}
\caption{Schematic drawing of the experimental setup:
M=monochromator, S=beam splitter, D$_{1,2}$=detectors,
C=coincidence counter, DAE=Data Acquisition Electronics. The two
detectors can be positioned at the same distance from S, and one
of them can be moved across this distance. The collimators are not
shown.}
\label{fig1}
\end{figure}

Let us start by considering the conceptual scheme of our
experiment, which is schematically represented in Fig.~\ref{fig1},
and is a massive particle analogue of the seminal optical
Hanbury-Brown and Twiss experiment \cite{sei}, which yielded the
first direct observation of the bunching effect in light beams and
is a direct consequence of the symmetric wave function of a
bosonic state. The semi-classical and quantum interpretations of
this experiment are very clearly discussed in classic textbooks
\cite{set,MandelWolf}.

In the present experiment, a monochromatic beam of thermal
neutrons is split into two  components by a pyrolytic graphite
single crystal that produces a transmitted and a reflected beam.
The intensity of each component is measured by a detector, and the
coincidence rate of the outcomes at the two detectors is recorded
by the coincidence counter as a function of their relative
distance from the splitting crystal. Of course, also accidental
coincidences occur in the apparatus and their rate must be
subtracted from the total rate. The average value of these rates
over a long enough period of observation provides the quantity to
be analyzed.

The nature of the emission of thermal neutrons in the source is
expected to be Poissonian, so that there is a small but finite
probability of having two neutrons within the detection time
interval of the apparatus. With reference to Fig.~\ref{fig1}, for
an average total rate $n$ of neutrons impinging on the splitter
crystal S and a total measuring time $T_0$, the predicted number
of coincidences $N_c$ measured at the two detectors can be
calculated from the joint probability that a neutron is
transmitted while a second neutron is reflected from S. By
neglecting the fermion antibunching, one readily obtains:
\begin{eqnarray}
N_{c}^{(\mathrm{id})} = \int_0^{T_0} dt \,  \tau_w n_t n_d =
\tau_w n_t n_d T_0 ,
\label{eq1}
\end{eqnarray}
where $n_t$ and $n_d = n - n_t$ are the average rates of the
transmitted and reflected beam respectively, that have been
assumed to be constant during the acquisition time $T_0$, and we
assumed a short enough coincidence window $\tau_w/2$, such that $n
\tau_w\ll 1$ (in our experiment $n \tau_w \simeq 10^{-2}\div
10^{-3}$). When the distances of the two detectors from the beam
splitter, $S_{\mathrm{D1}}$ and $S_{\mathrm{D2}}$, are different
enough, Eq.~(\ref{eq1}) gives the expected number of {\it random}
coincidences, since the measured coincidences are associated only
with the simultaneous detection of two particles emerging from the
splitter at different times.

Let us now qualitatively analyze the expected consequences of
antibunching. Again referring to an ideal experiment, let
$S_{\mathrm{D1}} = S_{\mathrm{D2}}$ and $\tau_c$ be the coherence
time of the neutron wave packet. It should be remarked that the
coherence time of the wave packet is defined by its energy
distribution and it is longer when the energy width is small.
Equation (\ref{eq1}) can be modified to yield the expected number
of {\it correlated} coincidences: observe that these are only
those due to two neutrons with different spins that emerge from
the beam splitter at the same time, because two neutrons with the
same spin, due to the Pauli exclusion principle, cannot impinge on
the beam splitter at the same time. If the incident beam is
spin-unpolarized, it is equally likely that a neutron pair will
either occur  in one of the three triplet states or in the singlet
state, i.e.\ the triplet states will occur 3/4 of the time. Thus,
the average number of coincidences expected for $S_{\mathrm{D1}} =
S_{\mathrm{D2}}$, as a consequence of fermion antisymmetry, is
reduced from that of Eq.~(\ref{eq1}) by the following quantity
\begin{eqnarray}
-\Delta N_{\rm fa} = - {1 \over 2}  \tau_c n_t n_d T_0 ,
\label{eq2}
\end{eqnarray}
where $-1/2=-3/4+1/4$, $-3/4$ being due to the antibunching of the
triplet states and $1/4$ to the bunching of the singlet state.
Such a depression of the coincidence rate involves a two-particle
state, and it is essential that both members of the pair be
detected. Since the relative directions of the two particles may
not be known in advance, one might think that something near a
$4\pi$ detector might generally be needed. But in the present
experiment the two-particle state to be tested is only that
emerging (within a small solid angle) from the collimator and the
monochromator, and the expression $\tau_c\, n_t \, n_d$ in
Eq.~(\ref{eq2}), is simply the rate of such emerging state. This
must be taken into account if one plans to perform a neutron-spin
test of the Bell inequality \cite{noi}.

Of course, in a real coincidence experiment, one must take
$\tau_w$ much longer than $\tau_c$, in order to account for
various instrumental effects that force one to broaden the
coincidence window. Actually, two particles arriving at the beam
splitter at the same time may be absorbed and recorded at the two
detectors within a rather long time interval $\tau_D$, because of
the finite thickness of the beam splitter and detectors and of
small differences in their speeds. Moreover, the finite detection
resolution $\tau_D <\tau_w$ tends to reduce the value of
$N_{c}^{(\mathrm{id})}$ in Eq.~(\ref{eq1}) by a factor
$\tau_w/\tau_D$, namely
\begin{eqnarray}
N_{c}^{(\infty)} = \frac{\tau_D}{\tau_w} N_{c}^{(\mathrm{id})} =
\tau_D n_t n_d T_0 ,
\label{eq:Ncinf}
\end{eqnarray}
making the use of intrinsically fast detectors highly desirable.
Clearly, as far as $\tau_D\gg\tau_c$, $\Delta N_{\rm fa}$ in
Eq.~(\ref{eq2}) remains unchanged. The experimental data require
therefore a careful analysis, as we shall discuss in detail in the
following.

In order to detect the expected fermion antibunching effect, we
have performed an optimized experiment based on the general scheme
of Fig.~\ref{fig1}. The main limitations of the experiment arise
from the random fluctuations, which can mask the {\it difference}
signal of Eq.~(\ref{eq2}). We assume that there are no accidental
coincidences due to non-random processes. Therefore the expected
random fluctuations are those due to the intrinsic statistics of
the number of measured coincidences. The expected root mean square
(rms) fluctuation of the total number of coincidences is given by:
\begin{eqnarray}
\Delta N_{c} = \sqrt{\tau_D n_t n_d T_0} .
\end{eqnarray}
In order to detect a signal it is necessary that the noise to
signal ratio,
\begin{eqnarray}
\frac{\Delta N_{c}}{\Delta N_{\rm fa}}= \frac{2\sqrt{\tau_D}}{
\tau_c \sqrt{n_t  n_d  T_0}}  ,
\label{eq3}
\end{eqnarray}
be smaller than unity. Looking at Eq.~(\ref{eq3}) we see that the
noise to signal ratio decreases when the coherence time of the
incoming beam is long, so that the experiment should be performed
by employing the most monochromatic available beam. We chose to
use the IN10 beam line \cite{in10}, which produces a monochromatic
beam by using an almost perfect Si(111) single crystal in the
backscattering configuration. This monochromator produces a flux
$n \simeq 2000\,$sec$^{-1}$, at an energy $E \simeq 2.08\,$meV
with an energy spread  $\Delta E \lesssim 0.13\, \mu$eV
(FWHM$=0.3\, \mu$eV). The coherence time of the incoming neutron
beam is therefore $\tau_c \gtrsim \hbar/2 \Delta E \simeq
2.5\,$ns. Considering that the neutron speed $v$ is about
$630\,$m/s, the neutron coherence length is larger than
1.5$\,\mu$m, a very small value. It is clear that both the beam
splitter and the detectors must be as thin as possible, in order
to reduce any additional spread of the signal. We therefore
employed a $0.3\,$mm thick graphite crystal as beam splitter. At
the wavelength of the present experiment the crystal has a good
reflectivity so that the transmitted and diffracted beam are of
the same order of magnitude. It should also be remarked that the
possible velocity difference between the two particles of a pair,
originating from the energy spread of the monochromator,
contributes a negligible difference in the corresponding time of
flight along the $40\,$cm path from the splitter to the detectors.

\begin{figure}
\begin{center}
\includegraphics[width=0.95\linewidth]{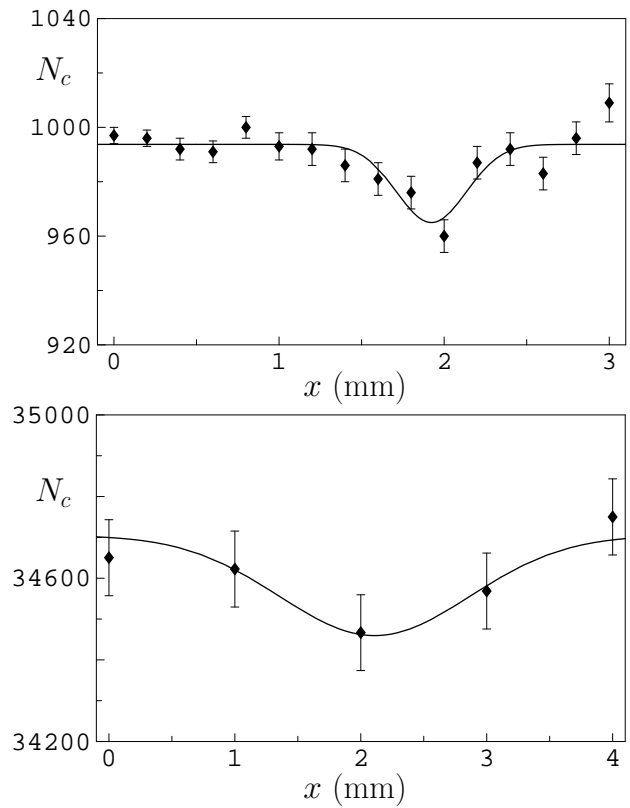}
\end{center}
\caption{Number of coincidences $N_c$ as a function of the path
difference $x=S_{\mathrm{D1}}-S_{\mathrm{D2}}$ of the detectors
from the splitter. Above: scintillator detectors, translation step
$0.2\,$mm; below: gas detector, translation step 1mm. The dip
appears at $x_0\simeq2\,$mm due to calibration. The parameters
$\tau_D, \tau_c, N_c^{(\infty)}$ and $x_0$, by substituting
$t=(x-x_0)/v$ in Eq.~(\ref{conteggi}), are determined by a best
fit (see text).}
\label{fig2}
\end{figure}

Two different detection systems were employed. The first one was
based on two squashed $^3$He $2\,$mm thick, $1.2\,$cm wide and
$10\,$cm high detectors. The second one was based on two
scintillator detectors having thickness $0.2\,$mm, width $1.5\,$cm
and height $5\,$cm. The scintillator was a  $^6$Li 98\% enriched
ZnS glass, directly coupled to a 5 cm diameter fast
photomultiplier. The shaping time was about $2\,\mu$s in the case
of $^3$He detectors, while it was $0.3\,\mu$s in the case of the
scintillators. The coincidences were measured within a time window
$\tau_w/2$ of $\pm 10\,\mu$s in the case of gas detectors and $\pm
0.8\,\mu$s for the scintillators.

Using this arrangement we have been able to perform two meaningful
determinations of the actual antibunching effect in the incoming
neutron beam. In order to do this, one detector was kept at a
fixed position, at a distance of $40\,$cm from the graphite
splitter along the diffracted beam, while the other detector was
scanned through the transmitted beam, at approximately the same
distance. For gas detectors, whose spatial resolution is of order
$2\,$mm, we used a coarse translation step of $1\,$mm, while for
scintillator detectors the translation step was $0.2\,$mm.

The data acquisition took several days; we therefore had to take
into account the effect of the incoming beam fluctuations. As can
be seen from Eq.~(\ref{eq:Ncinf}), such an effect is nonlinear and
directly related to the instantaneous value of the incoming beam
rate. Since the instantaneous rate cannot be measured with
adequate accuracy, one can perform a correction of the actual data
by assuming that the beam fluctuations are small. In such a case,
assuming that the acquisition time $T_0$ is much longer and the
detection window $\tau_w$ much shorter than the correlation time
of the noise and neglecting second order effects, Eqs.~(\ref{eq2})
and (\ref{eq:Ncinf}) yield
\begin{eqnarray}
N_{c} = \overline{N}_c \bigg[1 + {N_t - \overline{N}_t \over
\overline{N}_t} + {N_d - \overline{N}_d \over
\overline{N}_d}\bigg] \ ,
\label{eq4}
\end{eqnarray}
\noindent where $N_{c}$ is the actual number of coincidences,
$\overline{N}_c$ the number of coincidences detected in an (ideal)
experiment with a constant rate on both the transmitted and
incoming beam, and $N_t$ and $N_d$ are the actual numbers of
neutrons detected on the two beams. All the quantities in
Eq.~(\ref{eq4}) are functions of time, but vary on timescales that
are much slower than the time window $\tau_w$.

The experimental data collected with the two detecting systems
were corrected according to Eq.~(\ref{eq4}) and the results are
reported in Fig.~\ref{fig2}. In both cases a small dip is observed
in the number of coincidences detected as a function of the
relative distance of the two detectors from the beam splitter. We
attribute this small dip to the antibunching effect due to the
Fermion nature of the neutron.

It is interesting to perform a quantitative analysis, using the
experimentally observed width of the dip in order to get an
estimate of the coherence time of the incoming beam neutron wave
packet. As a spinoff, this will yield a consistency check of our
experimental results. Let us first consider the global response
function of our experimental arrangement. We assume that the
neutrons are diffracted within the thickness of the beam splitter
according to the secondary extinction law \cite{bacon} and are
absorbed by the detectors, within a few mean free paths inside the
absorbing medium. For simplicity, let us assume that the total
response function of the apparatus be (a normalized) Gaussian
\begin{eqnarray}
R(t) = \frac{1}{\sqrt{2\pi}\tau_D} \exp \left(-
\frac{t^2}{2\tau_D^2} \right)
\label{respfunct}
\end{eqnarray}
with a characteristic response time $\tau_D \gg \tau_c$, where
$\tau_c$ is the coherence time defined in Eq.~(\ref{eq2}). The
global response time $\tau_D$ of the apparatus will be obtained by
fitting the experimental data: it is however expected to be close
to the shaping time of the detectors ($2\,\mu$s for $^3$He
detectors, $0.3\,\mu$s for scintillators).

The neutron pair correlation function, describing the antibunching
effect, will also be taken to be Gaussian
\begin{eqnarray}
C(t) = 1 - {1 \over 2} \exp\left( -\frac{t^2}{2\tau_c^2}\right) ,
\label{eqC}
\end{eqnarray}
\noindent the factor 1/2=3/4-1/4 being due to the difference
between the triplet and the singlet contributions [see
Eq.~(\ref{eq2})]. The total number of counts is therefore given by
the convolution of these two functions
\begin{eqnarray}
\frac{N_{c}(t)}{N_c^{(\infty)}} & = & [R\ast C](t) =
\int ds \, R(s) C(t-s) \nonumber \\
& \simeq & 1 - \frac{1}{2} \frac{\tau_c}{\tau_D}\exp\left(-
\frac{t^2}{2\tau_D^2}\right) .
\label{conteggi}
\end{eqnarray}
This must be compared with the \emph{observed} number of
coincidences in Fig.~\ref{fig2}. Looking at the experimental data,
we see that in both cases there is a small but appreciable dip,
which is broader in the case of the experiment performed using the
(thicker) gas detectors, as expected. The above formula implies
that the width of the dip is $\tau_D$, its depth being $\tau_c/ 2
\tau_D=\Delta N_{\mathrm{fa}}/N_c^{(\infty)}$, in agreement with
Eqs.~(\ref{eq2}) and (\ref{eq:Ncinf}).

An accurate fit yields $\tau_D = 1.3\pm0.4\,\mu$s (with
$x_0=2.1\pm0.2\,$mm and $N_c^{(\infty)}=34720\pm44$) for $^3$He
detectors, and $\tau_D = 0.33\pm0.07\,\mu$s (with
$x_0=1.93\pm0.02\,$mm and $N_c^{(\infty)}=993.7\pm0.6$) for the
scintillators. Moreover,
\begin{eqnarray}\label{tauc}
\tau_c &=& 20\pm7\,\mathrm{ns} \quad \mbox{for $^3$He detectors},\\
\tau_c &=& 19\pm3\,\mathrm{ns} \quad \mbox{for scintillators},
\end{eqnarray}
both values being fully consistent with each other and with the
bound obtained by the energy spread of the beam ($\gtrsim
2.5\,$ns). The fitting curve is shown as a full line and is in
very good agreement with the data. Note that the value
$\tau_D\simeq 0.33\,\mu$s is in full accord with the nominal
shaping time of the scintillator ($0.3\,\mu$s), while the value
$\tau_D \simeq 1.3\,\mu$s is smaller than the nominal shaping time
of the $^3$He detectors ($2\,\mu$s): this can be understood by
remarking that $^3$He detectors tend to absorb neutrons in the
initial section. Finally, notice that one obtains results that are
consistent with the above ones also by performing a convolution
with more realistic (nongaussian) shape functions describing the
response function of the detectors and of the beam splitter.

It is useful to clarify in what sense this experiment performed
with neutrons is complementary to its photon \cite{sei,OuMandel}
and electron \cite{electron1,electron2,electron3} counterparts.
The first, obvious observation is that neutrons are fermions that
are not affected by Coulomb interaction, that plays, by contrast,
an important role in condensed matter systems. Second, neutrons
have very low phase-space densities, so that all the effects we
have brought to light are due to two-particle correlations, three
or more particle effects being completely negligible. This
experiment, providing a firm experimental evidence for the Pauli
exclusion principle, displaying its effects on free neutrons in
real space, has a very basic importance, because it is directly
related to the quantum mechanics of identical particles.

\acknowledgments We wish to thank Enzo Reali for his invaluable
technical collaboration. One of us (M.I.) is grateful to the
Istituto Nazionale di Fisica Nucleare (INFN) for financial
support.

\end{document}